\documentclass[a4paper,11pt]{article}
\usepackage{jheppub}

\newcommand{\Tr}{\operatorname{Tr}}
\renewcommand{\d}[2][]{\mathrm{d}^{#1}{#2}}
\renewcommand{\Re}[1]{\operatorname{Re}{#1}}
\renewcommand{\Im}[1]{\operatorname{Im}{#1}}

\allowdisplaybreaks

\title{
Duality and Supersymmetry Constraints on the Weak Gravity Conjecture
}

\author[a]{Gregory J. Loges,}
\author[b]{Toshifumi Noumi}
\author[a]{and Gary Shiu}

\affiliation[a]{Department of Physics, University of Wisconsin-Madison, Madison, WI 53706, USA}
\affiliation[b]{Department of Physics, Kobe University, Kobe 657-8501, Japan}

\emailAdd{gloges@wisc.edu}
\emailAdd{tnoumi@phys.sci.kobe-u.ac.jp}
\emailAdd{shiu@physics.wisc.edu}

\preprint{KOBE-COSMO-20-10, MAD-TH-20-03}

\abstract{
Positivity bounds coming from consistency of UV scattering amplitudes are not always sufficient to prove the weak gravity conjecture for theories beyond Einstein-Maxwell. Additional ingredients about the UV may be necessary to exclude those regions of parameter space which are na\"ively in conflict with the predictions of the weak gravity conjecture. In this paper we explore the consequences of imposing additional symmetries inherited from the UV theory on higher-derivative operators for Einstein-Maxwell-dilaton-axion theory. Using black hole thermodynamics, for a preserved SL($2,\mathbb{R}$) symmetry we find that the weak gravity conjecture then does follow from positivity bounds. For a preserved O($d,d;\mathbb{R}$) symmetry we find a simple condition on the two Wilson coefficients which ensures the positivity of corrections to the charge-to-mass ratio and that follows from the null energy condition alone. We find that imposing supersymmetry on top of either of these symmetries gives corrections which vanish identically, as expected for BPS states.
}

\begin{document}

\maketitle
\flushbottom

\section{Introduction}
The weak gravity conjecture (WGC) posits the existence of a state with charge larger than its mass, in appropriate units~\cite{ArkaniHamed:2006dz}.
In its mild from, it is enough to have the extremality bound of charged black holes be less stringent when corrections to the classical action are included \cite{Kats:2006xp,Cheung:2018cwt,Hamada:2018dde, Bellazzini:2019xts,Charles:2019qqt,Loges:2019jzs,Goon:2019faz,Cano:2019oma,Cano:2019ycn,Cremonini:2019wdk,Wei:2020bgk,Chen:2020rov,Cheung:2019cwi}.
While the mild form alone does not display the full predictive power for phenomenology\footnote{The power of the WGC lies with its constraints on low energy physics (most notably in constraining axion inflation models \cite{Brown:2015iha,Brown:2015lia,Montero:2015ofa, delaFuente:2014aca,Heidenreich:2015wga,Rudelius:2015xta,Junghans:2015hba,Bachlechner:2015qja, Cottrell:2016bty,
Hebecker:2016dsw,Hebecker:2017uix,Grimm:2019wtx,Heidenreich:2019bjd}). See also~\cite{Brennan:2017rbf,Palti:2019pca} for a review of the WGC and other swampland conjectures.}, 
when combined with additional UV properties such as modular invariance \cite{Heidenreich:2016aqi,Montero:2016tif,Aalsma:2019ryi} it can lead to stronger forms of the WGC such as the sublattice WGC \cite{Heidenreich:2016aqi,Montero:2016tif} and the tower WGC \cite{Andriolo:2018lvp}. For this reason, considerable efforts have been devoted toward a proof of the mild form of the WGC as a basis of the web of WGCs.

An important issue in this context is to identify consistency conditions necessary to demonstrate the conjecture. At low energies, corrections to the black hole extremality bound may be captured by higher-derivative operators, so that the mild form of the WGC follows if the effective couplings satisfy a certain inequality~\cite{Kats:2006xp}. Then, it is natural to expect that positivity bounds~\cite{Adams:2006sv} which follow from consistency of UV scattering amplitudes may play a crucial role in demonstrating the conjecture. It is indeed the case in Einstein-Maxwell theory under reasonable assumptions~\cite{Hamada:2018dde, Bellazzini:2019xts} (see also~\cite{Cheung:2014ega,  Andriolo:2018lvp, Chen:2019qvr} for attempts to use positivity bounds to constrain the charged particle spectrum at low energies).

However, there are some low-energy effective theories for which positivity bounds are not sufficient on their own to prove the positivity of corrections to the charge-to-mass ratio of extremal black holes. This occurs, for example, in the Einstein-Maxwell-dilaton theory, where the term
\begin{equation}
	\partial_\mu\phi\partial_\nu\phi F^{\mu\rho}F^\nu{}_\rho
\end{equation}
contributes to the charge-to-mass ratio in a way directly at odds with the WGC once positivity bounds are accounted for~\cite{Loges:2019jzs}. Of course, such a term never exists in isolation and presumably this negative contribution never dominates over other (positive) contributions. As discussed in~\cite{Loges:2019jzs}, for a few hand-picked choices of UV completion one can show that indeed this puzzling term is never problematic.

A similar scenario is present for the Einstein-axion-dilaton theory, where, as discussed recently in~\cite{Andriolo:2020lul}, there are regions of parameter space in which the axion weak gravity conjecture is violated, even when positivity bounds are taken into account. There, imposing extra structure on the UV theory, namely an SL($2,\mathbb{R}$) symmetry respected by the higher-derivative terms, greatly constrains the form of the corrections and ensures the axion WGC follows from the positivity bounds.

In this paper we consider the implications of imposing such additional structures on the UV theory for the WGC as applied to extremal black holes in Einstein-Maxwell-dilaton-axion (EMda) theory (see, e.g.,~\cite{Loges:2019jzs, Cano:2019ycn, Cano:2019oma} for previous discussions of the mild form of the WGC in the presence of a dilaton). Extra symmetries in the effective action which descend from the UV theory, when combined with either scattering positivity bounds or null energy conditions, are then strong enough to demonstrate the WGC for this system. In particular, we will work with an SL($2,\mathbb{R}$) symmetry and an O($d,d;\mathbb{R}$) symmetry: both can be present in the two-derivative EMda action and in 4D effective string actions these correspond to S- and T-duality, respectively~\cite{Schwarz:1993mg, Sen:1993sx, Sen:1994fa}. We also study implications of $N\geq2$ supersymmetry in these setups. We find that the puzzling terms mentioned earlier are helpful to make corrections to extremality identically zero even in the presence of nontrivial higher derivative operators, as expected for BPS states.

This paper is organized as follows: in section~\ref{sec:Symmetries} we recall the realizations of the SL($2,\mathbb{R}$) and O($d,d;\mathbb{R}$) symmetries for the two-derivative EMda action and impose these symmetries on the higher-derivative terms; in section~\ref{sec:LeadingOrderSoln} we present the leading order, dyonic solutions; in section~\ref{sec:Thermo} we use black hole thermodynamics to compute the corrections to the extremal charge-to-mass ratio, and we conclude in section~\ref{sec:Disc}. Throughout we use reduced Planck units: $8\pi G_\text{N}=1$.

\section{Symmetries of the Low-Energy Effective Action}
\label{sec:Symmetries}
Let us begin by recalling the two-derivative action for EMda theory: we will work with two U(1)s for simplicity. In discussing the two symmetries it is useful to go back-and-forth both between string and Einstein frame and between axion and 2-form field. Start in string frame,
\begin{equation}\label{eq:StringActionH2}
	I = \frac{1}{2}\int\d[4]{x}\,\sqrt{-g}\,e^{-2\phi}\Big[R + 4(\partial\phi)^2 - \frac{1}{12}H^2 - \frac{1}{2}F_a\cdot F_a\Big] \,,
\end{equation}
where $H = \d{B} - A_a\wedge F_a$ and the index $a=1,2$ is summed over. Here we have introduced the short-hand $G\cdot H\equiv G_{\mu\cdots\sigma}H^{\mu\cdots\sigma}$ and $G^2\equiv G\cdot G$ for antisymmetric tensors $G$ and $H$ of the same rank. Going to string frame via $g_{\mu\nu}\to e^{2\phi}g_{\mu\nu}$ gives
\begin{equation}\label{eq:EinsteinActionH2}
	I = \frac{1}{2}\int\d[4]{x}\,\sqrt{-g}\Big[ R - 2(\partial\phi)^2 - \frac{1}{12}e^{-4\phi}H^2 - \frac{1}{2}e^{-2\phi}F_a\cdot F_a\Big]\,.
\end{equation}
Dualizing to an axion is accomplished via
\begin{equation}
	I \supset \frac{1}{2}\int\Big[{-\frac{1}{2}}e^{-4\phi}{\star H}\wedge H - \theta\,\big(\d{H}+F_a\wedge F_a\big)\Big] \,.
\end{equation}
Integrating out $\theta$ reproduces~\eqref{eq:EinsteinActionH2}, while integrating out $H$ gives $H = e^{4\phi}{\star\d{\theta}}$ and
\begin{equation}\label{eq:EinsteinActionTheta2}
	I = \frac{1}{2}\int\d[4]{x}\,\sqrt{-g}\Big[ R - 2(\partial\phi)^2 - \frac{1}{2}e^{4\phi}(\partial\theta)^2 - \frac{1}{2}e^{-2\phi}F_a\cdot F_a + \frac{1}{2}\theta\,F_a\cdot\widetilde{F}_a \Big] \,,
\end{equation}
where $\widetilde{F}_{\mu\nu} = \frac{1}{2}\sqrt{-g}\,\epsilon_{\mu\nu\rho\sigma}F^{\rho\sigma}$ ($\epsilon_{0123}=-\epsilon^{0123}=+1$).

\subsection{$\operatorname{SL}(2,\mathbb{R})$}
The SL($2,\mathbb{R}$) symmetry is best discussed in Einstein frame, where by defining $\tau=\theta + ie^{-2\phi}$ and
\begin{equation}
	F_a^\pm = \frac{1}{2}\big(F_a\pm i \widetilde{F}_a\big) \,,
\end{equation}
the action becomes
\begin{equation}
	I = \frac{1}{2}\int\d[4]{x}\,\sqrt{-g}\Big[ R - \frac{\partial_\mu\tau\partial^\mu\overline{\tau}}{2(\Im\tau)^2} - \Im[\tau\,(F_a^-\cdot F_a^-)] \Big]\,.
\end{equation}
The SL($2,\mathbb{R}$) symmetry acts nonlinearly on the fields as\footnote{It is worth noting that the transformations of the fields are altered in the presence of higher-derivative terms in the action: for example, the $\alpha_{1111}$ term in~\eqref{eq:SL2Raction} induces
\begin{equation*}
	F_1^- \mapsto \big[ c\tau + d + 16(\Im\tau)^2\alpha_{1111}(F_1^{+2}) \big]F_1^-\,.
\end{equation*}
These ensure that the equations of motion are invariant under the SL($2,\mathbb{R}$) transformation at $\mathcal{O}(\alpha)$.}
\begin{align}\label{eq:SL2Rtransformations}
	g_{\mu\nu} &\mapsto g_{\mu\nu}\,, & \tau &\mapsto \frac{a\tau + b}{c\tau + d}\,, & F_a^- &\mapsto (c\tau + d)F_a^-\,,
\end{align}
where
\begin{equation}
	\begin{pmatrix}
		a & b\\
		c & d
	\end{pmatrix} \in \operatorname{SL}(2,\mathbb{R})\,,
\end{equation}
and is present at the level of the equations of motion. Electric and magnetic charges are defined by
\begin{equation}\label{eq:ChargeDefn}
	Q_a = \int\big(\Re\tau\;F+\Im\tau\;\widetilde{F}\big) = 2\int_\infty \Re(\tau F_a^-) \,, \quad P_a = \int F_a = 2\int_\infty \Re(F_a^-)\,,
\end{equation}
and using~\eqref{eq:SL2Rtransformations}, these transform under SL($2,\mathbb{R}$) according to
\begin{align}
	\begin{pmatrix}
		Q_a\\ P_a
	\end{pmatrix} &\mapsto \begin{pmatrix}
		a & b\\
		c & d
	\end{pmatrix}\begin{pmatrix}
		Q_a\\ P_a
	\end{pmatrix} \,.
\end{align}
We will make use of the rescaled charges $Q_a=4\pi q_a$ and $P_a=4\pi p_a$ as well.

A complete set of SL($2,\mathbb{R}$)-preserving\footnote{The SL($2,\mathbb{R}$) symmetry is known to be broken to SL($2,\mathbb{Z}$) due to non-perturbative effects. Imposing this less restrictive symmetry on the four-derivative action allows for a wider range of terms, e.g.~$r\big(j(\tau)\big)(\Im\tau)^2(F^{-2})(F^{+2})$ and $r\big(j(\tau)\big)(\Im\tau)^4G_4(\tau)(F^{+2})^2$, where $r\big(j(\tau)\big)$ is any rational function of the $j$--invariant and $G_4$ is the Eisenstein series of weight four. However, being non-perturbatively generated such terms will be highly suppressed.}, four-derivative operators may be written as
\begin{align}\label{eq:SL2Raction}
	\Delta I &= \int\d[4]{x}\,\sqrt{-g}\Big[ (\Im\tau)^2\,\alpha_{abcd}(F_a^-\cdot F_b^-)(F_c^+\cdot F_d^+) + (\Im\tau)^{-1}\,\alpha_{ab}(\partial_\mu\tau\partial_\nu\overline{\tau}\,F_a^{-\mu\rho}F_b^{+\nu}{}_\rho) \notag\\
	&\hspace{80pt} + (\Im\tau)^{-4}\big[\alpha_1(\partial_\mu\tau\partial^\mu\overline{\tau})^2 + \alpha_2|\partial_\mu\tau\partial^\mu\tau|^2\big] + \alpha_3 E^2 \Big] \,,
\end{align}
where $E^2=\mathrm{Riem}^2 - 4\,\mathrm{Ric}^2 + R^2$ is a total derivative in 4D. For the action to be real the Wilson coefficients must be real and have the following symmetries:
\begin{align}
	\alpha_{ab} &= \alpha_{ba}\,, & \alpha_{abcd} &= \alpha_{bacd} = \alpha_{abdc} = \alpha_{cdab} \,.
\end{align}
All-told there are 12 real parameters controlling the Wilson coefficients of equation~\eqref{eq:SL2Raction}. As mentioned in the introduction, without the imposed symmetry there are far more allowed terms and with regions of parameter space in conflict with the WGC that are not ruled out by positivity bounds. Note that the coefficient of the previously-noted term $\partial\phi\partial\phi FF$ is now related by the SL($2,\mathbb{R}$) symmetry to, among others, the coefficient of $(\partial\phi)^2(F^2)$.

\subsection{$\operatorname{O}(d,d;\mathbb{R})$}\label{sec:O(d,d)symmetry}
The O($d,d;\mathbb{R}$) symmetry is best discussed in string frame with the 3-form $H$. In reducing from $4+d$ to 4 dimensions on a torus, one finds a collection of KK scalars and U(1) gauge fields which collect themselves into a manifestly O($d,d;\mathbb{R}$)-invariant action\footnote{With $k$ gauge fields in $4+d$ dimensions the symmetry is enhanced to O($d,d+k;\mathbb{R}$). We will not consider this extension here.}. Upon reducing further to three, two or one dimension(s), the symmetry is O($d',d';\mathbb{R}$) with $d'=d+1,d+2,d+3$. When we refer to O($d,d;\mathbb{R}$) symmetry we have in mind this family of symmetries which appear when reducing on an arbitrary torus. The O($d,d;\mathbb{R}$)-symmetric four-derivative terms we discuss below are invariant under O($d',d';\mathbb{R}$) when reducing to even lower dimensions.

Using hats for $(4+d)$-dimensional indices and $\Phi$ for the $(4+d)$-dimensional dilaton, the decomposition
\begin{equation}
\begin{aligned}
	g_{\hat{\mu}\hat{\nu}} &= \begin{pmatrix}
		g_{\mu\nu} + A_\mu^{(1)p}G_{pq}A_\nu^{(1)q} & A_\mu^{(1)p}G_{pn}\\
		G_{mp}A_\nu^{(1)p} & G_{mn}
	\end{pmatrix} \,,\\
	B_{\hat{\mu}\hat{\nu}} &= \begin{pmatrix}
		B_{\mu\nu} - A_{[\mu}^{(1)m}A_{\nu]m}^{(2)} + A_\mu^{(1)m}B_{mn}A_\nu^{(1)n} & A_{\mu\,n}^{(2)} - B_{np}A_\mu^{(1)p}\\
		-A_{\nu\,m}^{(2)} + B_{mp}A_\nu^{(1)p} & B_{mn}
	\end{pmatrix} \,,\\
	\Phi &= 2\phi + \frac{1}{2}\log{\det{G_{mn}}}\,,\\
	\mathcal{A}_\nu^M &= \begin{pmatrix}
		A_\mu^{(1)m}\\
		A_{\mu\,m}^{(2)}
	\end{pmatrix}\,,
\end{aligned}
\end{equation}
and the matrices
\begin{equation}
\begin{aligned}
	\mathcal{H}_{MN} &= \begin{pmatrix}
		G_{mn} - B_{mp}G^{pq}B_{qn} & B_{mp}G^{pn}\\
		-G^{mp}B_{pn} & G^{mn}
	\end{pmatrix}\,,\\
	\eta^{MN} &= \begin{pmatrix}
		0 & \delta^m{}_n\\
		\delta_m{}^n & 0
	\end{pmatrix}
\end{aligned}
\end{equation}
bring the $(4+d)$-dimensional action,
\begin{equation}
	I = \frac{1}{2}\int\d[4+d]{x}\,\sqrt{-g}\,e^{-\Phi}\Big( R + \partial_{\hat{\mu}}\Phi\partial^{\hat{\mu}}\Phi - \frac{1}{12}H_{\hat{\mu}\hat{\nu}\hat{\rho}}H^{\hat{\mu}\hat{\nu}\hat{\rho}} \Big) \,,
\end{equation}
to the 4D form~\cite{Metsaev:1987zx, Meissner:1991ge, Schwarz:1992tn, Hohm:2015doa, Eloy:2020dko}
\begin{equation}
	I = \frac{1}{2}\int\d[4]{x}\,\sqrt{-g}\,e^{-2\phi}\Big[ R + 4(\partial\phi)^2 - \frac{1}{12}H^2 + \frac{1}{8}\Tr(\partial_\mu\mathcal{H}^{-1}\partial^\mu\mathcal{H}) - \frac{1}{4}\mathcal{F}_{\mu\nu}^M\mathcal{H}_{MN}\mathcal{F}^{\mu\nu\,N}\Big]\,,
\end{equation}
where $H = \d{B} -\frac{1}{2}\eta_{MN}\mathcal{A}^M\wedge\mathcal{F}^N$. This is manifestly invariant under $\mathcal{H}\to\Omega\mathcal{H}\Omega^\text{T}$ and $\mathcal{F}\to\Omega\mathcal{F}$ for $\Omega\in\mathrm{O}(d,d;\mathbb{R})$. We will consider the background with internal scalars $G_{mn}=\delta_{mn}$ and $B_{mn}=0$ so that
\begin{equation}
	\mathcal{H}_{MN} = \begin{pmatrix}
		\delta_{mn} & 0\\
		0 & \delta^{mn}
	\end{pmatrix} \equiv \delta_{MN} \,.
\end{equation}
There are only two independent four-derivative terms which respect the O($d',d';\mathbb{R}$) symmetry \cite{Eloy:2020dko}, which with the above choice for internal scalars read
\begin{equation}
\begin{aligned}
	\Delta I &= \int\d[4]{x}\,\sqrt{-g}\,e^{-2\phi}\Big\{ \alpha\Big[R_{\mu\nu\rho\sigma}R^{\mu\nu\rho\sigma} - \frac{1}{2}\delta_{MN}\big(R\mathcal{F}^M\mathcal{F}^N\big)\\
	&\qquad\qquad + \Big( \frac{1}{8}\delta_{MP}\delta_{NQ} - \frac{1}{2}\delta_{MQ}\delta_{NP} + \frac{1}{8}\eta_{MP}\eta_{NQ} \Big)\big(\mathcal{F}^M\mathcal{F}^N\mathcal{F}^P\mathcal{F}^Q\big) + \mathcal{O}(H^2)\Big]\\
	&\qquad + \beta\Big[ \frac{1}{4}\eta_{MN}\big(R\mathcal{F}^M\mathcal{F}^N\big) - \frac{1}{16}\eta_{MP}\delta_{NQ}\big(\mathcal{F}^M\cdot\mathcal{F}^N\big)\big(\mathcal{F}^P\cdot\mathcal{F}^Q\big) + \mathcal{O}(H) \Big] \Big\} \,,
\end{aligned}
\end{equation}
having introduced the notation
\begin{equation}
	\big(R\mathcal{F}^M\mathcal{F}^N\big) = R_{\mu\nu\rho\sigma}\mathcal{F}^{\mu\nu\,M}\mathcal{F}^{\rho\sigma\,N}\,, \qquad \big(\mathcal{F}^M\mathcal{F}^N\mathcal{F}^P\mathcal{F}^Q\big) = \mathcal{F}_{\mu\nu}^M\mathcal{F}^{\nu\rho\,N}\mathcal{F}_{\rho\sigma}^P\mathcal{F}^{\sigma\mu\,Q}\,.
\end{equation}
The bosonic and heterotic strings have $(\alpha,\beta)=(\frac{\alpha'}{8},0)$ and $(\alpha,\beta)=(\frac{\alpha'}{16},-\frac{\alpha'}{8})$, respectively. The $\beta$ term arises from dimesionally reducing a $(4+d)$-dimensional gravitational Chern-Simons term, and so should be accompanied by an $\mathcal{O}(\beta)$ correction to the Bianchi identity for $H$.

In going to Einstein frame and dualizing $H\to\theta$, one finds (using tree-level equations of motion)
\begin{equation}
\begin{aligned}
	I &= \frac{1}{2}\int\d[4]{x}\,\sqrt{-g}\Big\{ R - 2(\partial\phi)^2 - \frac{1}{2}e^{4\phi}(\partial\theta)^2 - \frac{1}{4}e^{-2\phi}\delta_{MN}\mathcal{F}^M\cdot\mathcal{F}^N\\
	&\quad + \frac{1}{4}\theta\,\eta_{MN}\mathcal{F}^M\cdot\widetilde{F}^N + 2\alpha\Big[ e^{-2\phi}E^2-\frac{1}{2}e^{-4\phi}\delta_{MN}\big(R\mathcal{F}^M\mathcal{F}^N\big)\\
	&\qquad + e^{-6\phi}\Big(\delta_{MN}\delta_{PQ} + \frac{1}{8}\delta_{MP}\delta_{NQ} - \frac{1}{2}\delta_{MQ}\delta_{NP} + \frac{1}{8}\eta_{MP}\eta_{NQ}\Big)\big(\mathcal{F}^M\mathcal{F}^N\mathcal{F}^P\mathcal{F}^Q\big)\\
	&\qquad + 2e^{-4\phi}\delta_{MN}\big(\partial\phi\partial\phi\mathcal{F}^M\mathcal{F}^N\big) - \frac{1}{8}e^{-6\phi}\big(\delta_{MN}\mathcal{F}^M\cdot\mathcal{F}^N\big)^2 + \mathcal{O}\big(\partial\theta^2\big) \Big]\\
	&\quad + 2\beta\Big[ \frac{1}{4}e^{-4\phi}\eta_{MN}\big(R\mathcal{F}^M\mathcal{F}^N\big) - e^{-4\phi}\eta_{MN}\big(\partial\phi\partial\phi\mathcal{F}^M\mathcal{F}^N\big)\\
	&\qquad + e^{-6\phi}\Big(\frac{1}{32}\eta_{MN}\delta_{PQ} - \frac{1}{16}\eta_{MP}\delta_{NQ}\Big)\big(\mathcal{F}^M\cdot\mathcal{F}^N\big)\big(\mathcal{F}^P\cdot\mathcal{F}^Q\big)\\
	&\qquad + \frac{1}{2}e^{-4\phi}(\partial\phi)^2\eta_{MN}\mathcal{F}^M\cdot\mathcal{F}^N + \mathcal{O}(\theta) \Big] \Big\} \,.
\end{aligned}
\end{equation}
We have omitted the higher-derivative terms involving the axion because they all vanish for the solution of section~\ref{sec:LeadingOrderSoln}.

\section{Leading-Order Solution}
\label{sec:LeadingOrderSoln}
The higher-derivative operators discussed in the previous section will induce corrections to extremal black holes; for the thermodynamic arguments of section~\ref{sec:Thermo}, we need only the uncorrected black hole solution in order to compute the leading corrections to the extremality condition.

We will consider dyonic black holes, where the solutions are regular even in the extremal limit, with constant axion for ease of calculation. This can be arranged via an appropriate SL($2,\mathbb{R}$) transformation, but does not represent the most general O($d,d;\mathbb{R}$) solution. Such solutions are given by
\begin{equation}\label{eq:solution}
\begin{aligned}
	\d{s^2} &= -f\,\d{t^2} + f^{-1}\d{r^2} + (r+\kappa_1)(r+\kappa_2)\,\big(\d{\vartheta^2} + \sin^2{\vartheta}\,\d{\varphi^2}\big) \,,\\
	f(r) &= \frac{r(r-2\xi)}{(r+\kappa_1)(r+\kappa_2)} \,,\\
	\theta &= 0 \,,\\
	e^{-2\phi} &= \frac{r+\kappa_1}{r+\kappa_2} \,,\\
	A_1 &= -\frac{q}{r+\kappa_1}\,\d{t} \,,\\
	A_2 &= -p\,\cos{\vartheta}\,\d{\varphi} \,.
\end{aligned}
\end{equation}
The physical charges are $Q=4\pi q$ and $P=4\pi p$, and the constants $\kappa_a>0$ are determined by
\begin{equation}
	q^2 = \kappa_1(\kappa_1+2\xi)\,,\qquad\qquad p^2 = \kappa_2(\kappa_2+2\xi) \,.
\end{equation}
The inner and outer horizons are located at $r=0$ and $r=2\xi$, respectively, so that extremality corresponds to $\xi\to0$. From the metric we may read off the mass, temperature and entropy:
\begin{equation}\label{eq:MTSfrommetric}
\begin{aligned}
	M &= 4\pi(\kappa_1+\kappa_2 + 2\xi) \,,\\
	T &= \frac{\xi}{2\pi(\kappa_1+2\xi)(\kappa_2+2\xi)} \,,\\
	S &= 8\pi^2(\kappa_1+2\xi)(\kappa_2+2\xi) \,.
\end{aligned}
\end{equation}
For large enough charges the black hole is large and the curvatures are small at the outer horizon, even at extremality. This ensures that the derivative expansion is under control.

\section{Higher-Derivative Corrections via Thermodynamics}
\label{sec:Thermo}
We will leverage black hole thermodynamics to compute corrections to the charge-to-mass ratio of extremal black holes, and so begin by recalling the key ingredients to this procedure. See~\cite{Reall:2019sah} for a more complete discussion of these ideas.

The full four-derivative action can be written as
\begin{equation}
	I = I_0 + \Delta I + I_\partial\,,
\end{equation}
where $I_0$ is the two-derivative action and $\Delta I$ all denotes higher-derivative terms with their corresponding Wilson coefficients, collectively denoted by $\alpha$. The contribution $I_\partial$ contains all boundary terms and is required for a well-defined variational principle: the details of its form will not be relevant to our discussion. The Gibbons-Hawking-York term contributes $\frac{M}{2T}$ to the action~\cite{Gibbons:1976ue}, and the boundary terms associated with the other terms in the bulk action vanish in the infinite-volume limit.

One may evaluate the Euclidean action to find the free energy $G$ in the grand canonical ensemble, given by
\begin{equation}
\begin{aligned}
	T\,I_\text{E} = G &\equiv M - TS - Q_a\Phi_a \,,\\
	\d{G} &= -S\,\d{T} - Q_a\,\d{\Phi_a} + \Psi_a\,\d{P_a}\,,
\end{aligned}
\end{equation}
with $\Phi_a = (-A_t^a)|_{r=2\xi}-(-A_t^a)|_{r=\infty}$ the electric potentials at the outer horizon and $\Psi_a$ the analogous magnetic potential. Via straightforward thermodynamic manipulations one may find the mass as a function of the charges and temperature in the canonical ensemble.

The strength of this approach lies in there being no need to find solutions to the perturbed equations of motion. Namely, the Euclidean action may be reliably evaluated to $\mathcal{O}(\alpha)$ in the grand canonical ensemble using only the leading-order solution\footnote{This fails in 3D even for pure Einstein gravity, where the fall-off of boundary counter-terms is less than those in $D\geq4$.}:
\begin{equation}
	I_\text{E}[X(T,\Phi,P,\alpha)] = I_\text{E}[X(T,\Phi,P,0)] + \mathcal{O}(\alpha^2)\,.
\end{equation}
The dynamical fields are collectively denoted by $X$, and may even include fields which vanish at $\mathcal{O}(\alpha^0)$.

\subsection{Leading-Order Thermodynamics}
For the solution of~\eqref{eq:solution}, we discuss briefly the determination of the black hole's thermodynamic properties via the Euclidean action. This will give the leading behavior, on top of which we compute corrections due to the four-derivative terms. Evaluating the two-derivative Euclidean action for \eqref{eq:solution} leads to
\begin{equation}
	G(T,\Phi,P) = \frac{1-\Phi^2}{2T} + \frac{P^2T}{2(1-\Phi^2)}\,,
\end{equation}
from which it follows that in the grand canonical and canonical ensembles we have
\begin{equation}
\begin{aligned}
	M(T,\Phi,P) &= \frac{1}{T}\left(1-\frac{\Phi^2P^2T^2}{(1-\Phi^2)^2}\right)\,, & M(T,Q,P) &= Q\,w_q + P\,w_p+\frac{QPT}{w_qw_p}\,,\\
	Q(T,\Phi,P) &= \frac{\Phi}{T}\left(1 - \frac{P^2T^2}{(1-\Phi^2)^2}\right)\,, & \Phi(T,Q,P) &= w_q\,,\\
	\Psi(T,\Phi,P) &= \frac{PT}{1-\Phi^2}\,, & \Psi(T,Q,P) &= w_p\,,\\
	S(T,\Phi,P) &= \frac{1-\Phi^2}{2T^2}\left(1-\frac{P^2T^2}{(1-\Phi^2)^2}\right)\,, & S(T,Q,P) &= \frac{QP}{2w_qw_p}\,.
\end{aligned}
\end{equation}
We have used $w_q=w(QT,PT)$ and $w_p=w(PT,QT)$ in the canonical ensemble, with $w(y,z)$ being the root of the quintic
\begin{equation}
	f(x) = (x-y)(x^2-1)^2 - z^2x
\end{equation}
with a small $y,z>0$ expansion which begins
\begin{equation}
	w(y,z) = 1 - \frac{z}{2} - \frac{z}{8}(2y+z) + \cdots\,.
\end{equation}
This root is identified by its giving a positive mass which remains finite for $T\to0$. The other branches of solutions correspond to different signs for $q,p$ or to thermodynamically unstable configurations. The expressions above exactly match those found by reading off from the metric in equation~\eqref{eq:MTSfrommetric} upon eliminating $\kappa_1$, $\kappa_2$ and $\xi$ in favor of $Q$, $P$ and $T$.

\subsection{$\operatorname{SL}(2,\mathbb{R})$}
\begin{figure}[t]
	\centering
	\includegraphics[width=\textwidth]{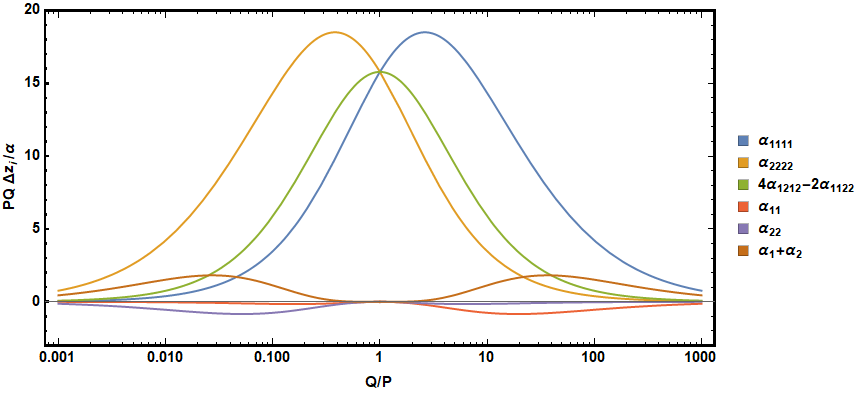}
	\caption{(Rescaled) individual contributions to $z_\text{ext}$ from each terms in the SL($2,\mathbb{R}$)-invariant action. The preserved electromagnetic duality ($Q\leftrightarrow P$ and $1\leftrightarrow2$) is clear from the symmetry about the $Q=P$ line.}
	\label{fig:SL2R_delta_z}
\end{figure}

For the sake of example we present briefly the results for the $\alpha_{1111}$ term in~\eqref{eq:SL2Raction} (other corrections have a similar form). In the grand canonical ensemble, we find
\begin{align}
	M(T,\Phi,P) &= \frac{1}{T}\left(1 - \frac{\Phi^2P^2T^2}{(1-\Phi^2)^2}\right) \notag\\
	&\quad + \frac{64\pi^2\alpha_{1111}\Phi^4T}{5(1-\Phi^2)^2}\bigg[ 2(2-\Phi^2)\,{}_2F_1\big(1,1;6;\tfrac{P^2T^2-\Phi^2(1-\Phi^2)^2}{(1-\Phi^2)^3}\big)\\
	&\quad + \left(-\frac{\Phi^2(1-\Phi^2)}{3(1-\Phi^2)^2} + \frac{P^2T^2(1+2\Phi^2)}{3(1-\Phi^2)^3}\right){}_2F_1\big(2,2;7;\tfrac{P^2T^2-\Phi^2(1-\Phi^2)^2}{(1-\Phi^2)^3}\big) \bigg]\,, \notag\\
	Q(T,\Phi,P) &= \frac{\Phi}{T}\left(1-\frac{P^2T^2}{(1-\Phi^2)^2}\right) \notag\\*
	&\quad + \frac{64\pi^2\alpha_{1111}\Phi^3T}{5(1-\Phi^2)^2}\bigg[ 2(2-\Phi^2)\,{}_2F_1\big(1,1;6;\tfrac{P^2T^2-\Phi^2(1-\Phi^2)^2}{(1-\Phi^2)^3}\big)\\*
	&\quad + \left(\frac{\Phi^2[3P^2T^2-(1-\Phi^2)^2]}{3(1-\Phi^2)^3}\right){}_2F_1\big(2,2;7;\tfrac{P^2T^2-\Phi^2(1-\Phi^2)^2}{(1-\Phi^2)^3}\big) \bigg] \,, \notag\\
	\Psi(T,\Phi,P) &= \frac{PT}{1-\Phi^2} + \frac{128\pi^2\alpha_{1111}P\Phi^4T^3}{5(1-\Phi^2)^4}\,{}_2F_1\big(2,2;7;\tfrac{P^2T^2-\Phi^2(1-\Phi^2)^2}{(1-\Phi^2)^3}\big)\,,\\
	S(T,\Phi,P) &= \frac{1-\Phi^2}{2T^2}\left(1-\frac{P^2T^2}{(1-\Phi^2)^2}\right) \notag\\
	&\quad + \frac{64\pi^2\alpha_{1111}T^4}{5(1-\Phi^2)^4}\bigg[(1-\Phi^2)^3\,{}_2F_1\big(1,1;6;\tfrac{P^2T^2-\Phi^2(1-\Phi^2)^2}{(1-\Phi^2)^3}\big)\\
	&\quad + \frac{P^2T^2}{3}\,{}_2F_1\big(2,2;7;\tfrac{P^2T^2-\Phi^2(1-\Phi^2)^2}{(1-\Phi^2)^3}\big)\bigg]\,.\notag
\end{align}
In the canonical ensemble, we find
\begin{align}
	M(T,Q,P) &= Q\,w_q + P\,w_p + \frac{QP\,T}{w_qw_p} - \frac{128\pi^2\alpha_{1111}w_q^3T}{5(1-w_q^2)[4w_q^2-\frac{QT}{w_q}(1+3w_q^2)]} \notag\\
	&\qquad \times\bigg[ \left[ 2w_q(2-w_q^2) - QT\left(3-w_q^2\right) \right]\,{}_2F_1\big(1,1;6;1-\tfrac{QT}{w_q(1-w_q^2)}\big)\\
	&\qquad\qquad + QT\,{}_2F_1\big(1,2;6;1-\tfrac{QT}{w_q(1-w_q^2)}\big) \bigg] \,, \notag\\
	\Phi(T,Q,P) &= w_q + \frac{64\pi^2\alpha_{1111}w_q^3T^2}{5(1-w_q^2)[4w_q^2-\frac{QT}{w_q}(1+3w_q^2)]} \notag\\
	&\qquad \times\bigg[ 2(2-w_q^2)\,{}_2F_1\big(1,1;6;1-\tfrac{QT}{w_q(1-w_q^2)}\big)\\
	&\qquad\qquad + \frac{w_q(2w_q-3QT)}{3(1-w_q^2)}\,{}_2F_1\big(2,2;7;1-\tfrac{QT}{w_q(1-w_q^2)}\big) \bigg]\,, \notag\\
	\Psi(T,Q,P) &= w_p + \frac{128\pi^2\alpha_{1111}Pw_q^3T^3}{5(1-w_q^2)^3(1-w_q^2+\frac{QT}{w_q})[4w_q^2-\tfrac{QT}{w_q}(1+3w_q^2)]} \notag\\
	&\qquad \times\bigg[ (1-w_q^2)\big[ 2w_q(2-w_q^2)-5QT \big]\,{}_2F_1\big(1,1;6;1-\tfrac{QT}{w_q(1-w_q^2)}\big)\\
	&\qquad\qquad+ QT(1-3w_q^2)\,{}_2F_1\big(1,2;6;1-\tfrac{QT}{w_q(1-w_q^2)}\big) \bigg] \,,\notag\\
	S(T,Q,P) &= \frac{QP}{2w_qw_p} - \frac{64\pi^2\alpha_{1111}w_q^3}{5(1-w_q^2)[4w_q^2-\frac{QT}{w_q}(1+3w_q^2)]} \notag\\
	&\qquad \times\bigg[ (1-w_q^2)(8w_q-5QT)\,{}_2F_1\big(1,1;6;1-\tfrac{QT}{w_q(1-w_q^2)}\big)\\
	&\qquad\qquad + 2QT\,{}_2F_1\big(1,2;6;1-\tfrac{QT}{w_q(1-w_q^2)}\big) \bigg]\,. \notag
\end{align}
Taking the $T\to0$ limit at fixed charges gives the corrected extremal charge-to-mass ratio:
\begin{equation}\label{eq:zextSL2R}
\begin{aligned}
	z_\text{ext} &= 1 + \frac{1}{5p(p+q)}\Big[ (\alpha_{1111})\,{}_2F_1(1,1;6;1-\tfrac{Q}{P}) + (\alpha_{2222}){}_2F_1(1,5;6;1-\tfrac{Q}{P})\\
	&\qquad + (4\alpha_{1212}-2\alpha_{1122})\,{}_2F_1(1,3;6;1-\tfrac{Q}{P})\\
	&\qquad - \frac{\alpha_{11}}{84}\big(1-\tfrac{Q}{P}\big)^2{}_2F_1(3,3;8;1-\tfrac{Q}{P}) - \frac{\alpha_{22}}{84}\big(1-\tfrac{Q}{P}\big)^2{}_2F_1(3,5;8;1-\tfrac{Q}{P})\\
	&\qquad + \frac{\alpha_1+\alpha_2}{126}\big(1-\tfrac{Q}{P}\big)^4{}_2F_1(5,5;10;1-\tfrac{Q}{P}) \Big] \,.
\end{aligned}
\end{equation}
Each contribution is shown in figure~\ref{fig:SL2R_delta_z} separately. There the invariance of $z_\text{ext}$ under the preserved electromagnetic duality ($Q\leftrightarrow P$ and $1\leftrightarrow2$) is more clearly seen.

One may obtain positivity bounds on the coefficients present in $z_\text{ext}$ by considering scattering amplitudes around the background $g_{\mu\nu}=\eta_{\mu\nu}$, $\tau=i$ and $A_a=0$, such as appears in the asymptotic region of the black holes considered here. For $\mathcal{M}(s,t=0)$ a crossing-symmetric forward amplitude, we have~\cite{Adams:2006sv}
\begin{equation}
\begin{aligned}
	{}[s^2]\mathcal{M}(s,t=0) &= \oint_\mathcal{C}\frac{\d{s}}{2\pi i}\,\frac{\mathcal{M}(s,t=0)}{s^3} = \left(\int_{-\infty}^{-s_0} + \int_{s_0}^\infty\right)\frac{\d{s}}{2\pi i}\frac{\operatorname{Disc}{\mathcal{M}(s,t=0)}}{s^3}\\
	&= \frac{2}{\pi}\int_{s_0}^\infty\d{s}\,\frac{\Im{\mathcal{M}(s,t=0)}}{s^3} \geq 0 \,.
\end{aligned}
\end{equation}
The contour $\mathcal{C}$ encircles the origin and is deformed to two integrations along cuts beginning at $\pm s_0$: the contributions at infinity are dropped, having assumed the Froissart bound. By the optical theorem the imaginary part of the crossing-symmetric amplitude $\mathcal{M}$ is positive, showing that the coefficient of $s^2$ in $\mathcal{M}$ is also positive.

Assuming gravitational effects are subdominant (see appendix~\ref{sec:grav_positivity} for more discussion on this point), one has the following forward scattering amplitudes:
\begin{equation}
\begin{aligned}
	\mathcal{M}(A_a^\pm A_b^\pm A_c^\mp A_d^\mp) &= -\delta_{ab}\delta_{cd}\,s + 4\alpha_{abcd}\,s^2 \,,\\
	\mathcal{M}(A_a^\pm\phi A_b^\mp\phi) &= -\alpha_{ab}\,s^2\,,\\
	\mathcal{M}(\phi\phi\phi\phi) = \mathcal{M}(\theta\theta\theta\theta) &= 16(\alpha_1+\alpha_2)s^2 \,.
\end{aligned}
\end{equation}
From these we may read off that $\alpha_1+\alpha_2\geq0$ and $\alpha_{11},\alpha_{22}\leq0$, so that their corresponding contributions to $z_\text{ext}$ are each positive. For the four-photon amplitudes, the crossing-symmetric combinations
\begin{equation}
	\sum_{a,b,c,d}u_a v_b u_c v_d\;\alpha_{abcd}
\end{equation}
must be positive for all real $u_a,v_a$. In particular, choosing $u=(1,0)$, $v=(0,1)$ shows that $\alpha_{1212}\geq0$, and $u=(1,x)$, $v=(1,-x)$ gives
\begin{equation}
	\alpha_{1111} - 2x^2\alpha_{1122} + x^4\alpha_{2222} \geq 0
\end{equation}
for all real $x$. The choice
\begin{equation}
	x^4 = \frac{{}_2F_1\big(1,5;6;1-\frac{Q}{P}\big)}{{}_2F_1\big(1,1;6;1-\frac{Q}{P}\big)} > 0
\end{equation}
is enough to conclude that $z_\text{ext}\geq1$ in equation~\eqref{eq:zextSL2R}.

\subsection{$\operatorname{O}(d,d;\mathbb{R})$}
For those theories with a preserved O($d,d;\mathbb{R}$) symmetry the gravitational four-derivative terms are always of the same order as those involving the gauge fields. Without a hierarchy among the Wilson coefficients, we do not have positivity bounds from scattering amplitudes at our disposal. However, since the O($d,d;\mathbb{R}$) symmetry is far more constraining than SL($2,\mathbb{R}$), the four-derivative terms are controlled by only two undetermined coefficients and the interplay between positive and negative contributions to $z_\text{ext}$ is nearly fixed.

It is necessary to make connection between the gauge fields discussed in section~\ref{sec:O(d,d)symmetry} and the gauge fields solution of section~\ref{sec:LeadingOrderSoln}. That is, we should identify $A_{1,2}$ with components of $\mathcal{A}^M$ in such a way that
\begin{equation}
	\delta_{MN}\mathcal{F}^M\cdot\mathcal{F}^N = 2F_a\cdot F_a \qquad\text{and}\qquad \eta_{MN}\mathcal{F}^M\cdot\widetilde{\mathcal{F}}^N = \pm 2F_a\cdot\widetilde{F}_a \,.
\end{equation}
(The sign of $F_a\cdot \widetilde{F}_a$ may be absorbed into $\theta\to-\theta$.) Up to O($d,d;\mathbb{R}$) transformations this is accomplished with
\begin{equation}\label{eq:Achoices}
	\mathcal{A}^M = \begin{pmatrix}
		A_1\\
		A_2\\
		0\\
		\vdots\\ \hline
		\pm A_1\\
		\pm A_2\\
		0\\
		\vdots
	\end{pmatrix} \,.
\end{equation}

The extremal charge-to-mass ratio may be written as
\begin{equation}
	z_\text{ext} = 1 + \alpha\Delta z_\alpha + \beta\Delta z_\beta \,,
\end{equation}
where
\begin{align}
&\begin{aligned}
	\Delta z_\alpha &= \Delta z_{E^2} - \Delta z_{RF_1F_1} - \Delta z_{RF_2F_2} + 4\Delta z_{\partial\phi\partial\phi F_1F_1}\\
	&\qquad\qquad - \frac{1}{2}\Delta z_{(F_1^2)^2} - \Delta z_{(F_1^2)(F_2^2)} - \frac{1}{2}\Delta z_{(F_2^2)^2} + 3\Delta z_{F_1^4} + 3\Delta z_{F_2^4}\,,
\end{aligned}\\
&\begin{aligned}
	\Delta z_\beta &= \pm\Big(\frac{1}{2}\Delta z_{RF_1F_1} + \frac{1}{2}\Delta z_{RF_2F_2} - 2\Delta z_{\partial\phi\partial\phi F_1F_1} + \Delta z_{(\partial\phi)^2(F_1^2)} + \Delta z_{(\partial\phi)^2(F_2^2)}\\
	&\qquad\qquad -\frac{1}{8}\Delta z_{(F_1^2)^2} + \frac{1}{4}\Delta z_{(F_1^2)(F_2^2)} - \frac{1}{8}\Delta z_{(F_2^2)^2}\Big)
\end{aligned}
\end{align}
(any term not present vanishes). The individual terms are (see figure~\ref{fig:O(d,d)_delta_z})

\begin{figure}[t]
	\centering
	\includegraphics[width=\textwidth]{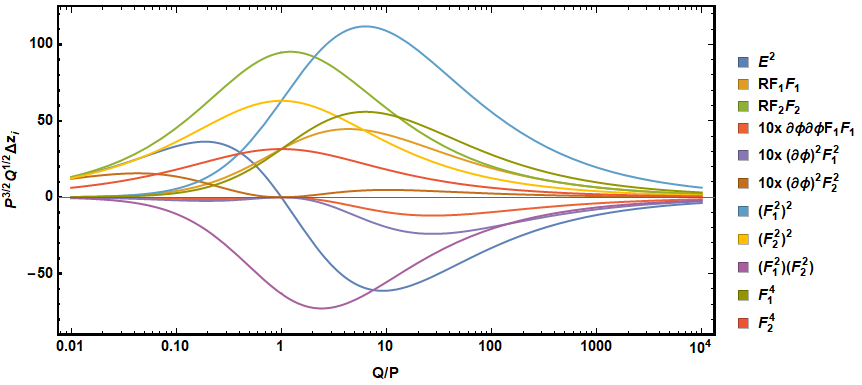}
	\caption{(Rescaled) individual contributions $\Delta z_i$ to $\Delta z_\alpha$ and $\Delta z_\beta$.}
	\label{fig:O(d,d)_delta_z}
\end{figure}

\begin{align}
	\Delta z_{E^2} &= -\frac{1}{1260p^6(p+q)}\Big[ 1120p^4q\,{}_2F_1\big(1,6;10;1-\tfrac{q}{p}\big)\\
	&\qquad + 280p^3q(p+q)\,{}_2F_1\big(2,6;10;1-\tfrac{q}{p}\big) \notag\\
	&\qquad - 2p\big(6p^4+133p^3q+337p^2q^2+67pq^3-3q^4\big)\,{}_2F_1\big(3,6;10;1-\tfrac{q}{p}\big) \notag\\
	&\qquad - q\big(41p^4+363p^3q+217p^2q^2-23pq^3+2q^4\big)\,{}_2F_1\big(4,6;10;1-\tfrac{q}{p}\big) \Big]\,, \notag\\
	\Delta z_{RF_1F_1} &= \frac{q}{35p^4(p+q)}\Big[ 20p^2\,{}_2F_1\big(1,4;8;1-\tfrac{q}{p}\big) - 4pq\,{}_2F_1\big(3,4;8;1-\tfrac{q}{p}\big)\\
	&\hspace{100pt} - q(p+q)\,{}_2F_1\big(4,4;8;1-\tfrac{q}{p}\big) \Big]\,, \notag\\
	\Delta z_{RF_2F_2} &= \frac{1}{210p^3(p+q)}\Big[ 24p(p+8q)\,{}_2F_1\big(1,6;8;1-\tfrac{q}{p}\big)\\
	&\hspace{100pt} + 2q(17p+q)\,{}_2F_1\big(2,6;8;1-\tfrac{q}{p}\big) \Big] \,,\notag\\
	\Delta z_{\partial\phi\partial\phi F_1F_1} &= -\frac{q(p-q)^2}{420p^4(p+q)}\,{}_2F_1\big(3,4;8;1-\tfrac{q}{p}\big)\,,\\
	\Delta z_{(\partial\phi)^2(F_1^2)} &= -\frac{q(p-q)^2}{210p^4(p+q)}\,{}_2F_1\big(3,4;8;1-\tfrac{q}{p}\big)\,,\\
	\Delta z_{(\partial\phi)^2(F_2^2)} &= \frac{q(p-q)^2}{210p^4(p+q)}\,{}_2F_1\big(3,6;8;1-\tfrac{q}{p}\big)\,,\\
	\Delta z_{(F_1^2)^2} &= \frac{4q}{5p^2(p+q)}\,{}_2F_1\big(1,2;6;1-\tfrac{q}{p}\big)\,,\\
	\Delta z_{(F_2^2)^2} &= \frac{4}{5p(p+q)}\,,\\
	\Delta z_{(F_1^2)(F_2^2)} &= -\frac{4q}{5p^2(p+q)}\,{}_2F_1\big(1,4;6;1-\tfrac{q}{p}\big)\,,\\
	\Delta z_{F_1^4} &= \frac{2q}{5p^2(p+q)}\,{}_2F_1\big(1,2;6;1-\tfrac{q}{p}\big)\,,\\
	\Delta z_{F_2^4} &= \frac{2}{5p(p+q)}\,.
\end{align}
However, with the O($d,d;\mathbb{R}$) symmetry present the combinations $\Delta z_\alpha$ and $\Delta z_\beta$ are quite simple:
\begin{equation}
\begin{aligned}
	\Delta z_\alpha &= \frac{4}{5p(p+q)}\,,\quad & \Delta z_\beta &= \pm\frac{2}{5p(p+q)}\,.
\end{aligned}
\end{equation}
Requiring $z_\text{ext}\geq1$ then amounts to $2\alpha\pm\beta\geq0$, or $2\alpha\geq|\beta|$. This WGC bound implies that the coefficient of the Gauss-Bonnet term, $\alpha$, is positive, in agreement with other considerations such as string theory examples and entropy arguments. As we now show, the condition $2\alpha\geq|\beta|$ follows from the null energy condition alone.

On any background and for any null vector $k^\mu$, the null energy condition requires
\begin{equation}
	T_{\mu\nu}k^\mu k^\nu \geq 0 \,,
\end{equation}
where the stress tensor $T_{\mu\nu}$ has contributions from both the two- and four-derivative terms in the action:
\begin{equation}
	T_{\mu\nu} = T_{\mu\nu}^{(2)} + T_{\mu\nu}^{(4)} \,.
\end{equation}
The $T_{\mu\nu}^{(4)}$ terms are explicitly of order $\alpha$, but there are also implicit $\mathcal{O}(\alpha)$ corrections from evaluating $T_{\mu\nu}^{(2)}$ on the corrected background.

For our purposes it will suffice to work with the spherically-symmetric background with $p=q$, so that $\phi,\theta=\mathcal{O}(\alpha)$ and many terms drop out of $T_{\mu\nu}^{(2)}$ and $T_{\mu\nu}^{(4)}$. With the choice
\begin{equation}
	k^\mu = \big\langle \sqrt{-g^{tt}},\sqrt{g^{rr}},0,0\big\rangle \,,
\end{equation}
the leading contribution vanishes:
\begin{equation}
\begin{aligned}
	T_{\mu\nu}^{(2)}k^\mu k^\nu &= e^{-2\phi}\delta_{ab}\Big(F_{a\,\mu\rho}F_{b\,\nu}{}^\rho - \frac{1}{4}g_{\mu\nu}F_a\cdot F_b\Big) k^\mu k^\nu + 2\Big[\partial_\mu\phi\partial_\nu\phi - \frac{1}{2}g_{\mu\nu}(\partial\phi)^2\Big]k^\mu k^\nu\\
	&\qquad\qquad + \frac{1}{2}e^{4\phi}\Big[\partial_\mu\theta\partial_\nu\theta - \frac{1}{2}g_{\mu\nu}(\partial\theta)^2\Big]k^\mu k^\nu\\
	&= e^{-2\phi}\delta_{ab}\big({-g^{tt}}F_{a\,t\rho}F_{b\,t}{}^\rho + g^{rr}F_{a\,r\rho}F_{b\,r}{}^\rho + 2\sqrt{-g^{tt}g^{rr}}F_{a\,t\rho}F_{b\,r}{}^\rho\big) + \mathcal{O}(\alpha^2)\\
	&= 0 + \mathcal{O}(\alpha^2)\,.
\end{aligned}
\end{equation}
The last equality follows from the spherical symmetry of the corrected solution, for which $g_{\mu\nu}$ remains diagonal and only $F_{a\,tr}$ and $F_{a\,\vartheta\varphi}$ are potentially nonzero. Thus on this background only the explicit $\mathcal{O}(\alpha)$ terms of $T_{\mu\nu}^{(4)}$ can possibly give nonzero contribution to $T_{\mu\nu}k^\mu k^\nu$:
\begin{equation}
\begin{aligned}
	T_{\mu\nu}^{(4)} &= \alpha\Big[ {-T_{\mu\nu}^{RFF}} - \frac{1}{2}T_{\mu\nu}^{(F_1^2)^2} - T_{\mu\nu}^{(F_1^2)(F_2^2)} - \frac{1}{2}T_{\mu\nu}^{(F_2^2)^2} + 3T_{\mu\nu}^{F^4} \Big]\\
	&\qquad\qquad \pm \beta\Big[ \frac{1}{2}T_{\mu\nu}^{RFF} - \frac{1}{8}F_{\mu\nu}^{(F_1^2)^2} + \frac{1}{4}T_{\mu\nu}^{(F_1^2)(F_2^2)} - \frac{1}{8}T_{\mu\nu}^{(F_2^2)^2} \Big]\,,\\
	T_{\mu\nu}^{RFF} &= g_{\mu\nu}(\delta_{ab}RF_aF_b) - 6R_{\mu\alpha\beta\gamma}\delta_{ab}F_{a\,\nu}{}^\alpha F_b^{\beta\gamma} - 4\nabla^\beta\nabla^\alpha(F_{\mu\alpha}F_{\nu\beta})\,,\\
	T_{\mu\nu}^{F^4} &= \sum_{a=1,2}\big(g_{\mu\nu}F_aF_aF_aF_a-8F_{a\,\mu}{}^\sigma F_{a\,\alpha\beta}F_a^{\beta\gamma}F_{a\,\gamma\nu}\big)\,,\\
	T_{\mu\nu}^{(F_a^2)^2} &= g_{\mu\nu}(F_a^2)^2 - 8F_{a\,\mu}{}^\alpha F_{a\,\nu\alpha}(F_a^2)\,,\\
	T_{\mu\nu}^{(F_1^2)(F_2^2)} &= g_{\mu\nu}(F_1^2)(F_2^2) - 4F_{1\,\mu}{}^\alpha F_{1\,\nu\alpha}(F_2^2) - 4F_{2\,\mu}{}^\alpha F_{2\,\nu\alpha}(F_1^2)\,.
\end{aligned}
\end{equation}
These terms we may simply evaluate on the leading-order solution of equation~\eqref{eq:solution} with $p=q$ and $\kappa_2=\kappa_1$. In contracting with $k^\mu$ the only contribution which does not vanish is from the last term in $T_{\mu\nu}^{RFF}$, which leads to
\begin{equation}
	T_{\mu\nu}k^\mu k^\nu = (2\alpha\mp\beta)\frac{8q^2r(r-2\xi)}{(r+\kappa_1)^8} \geq 0\,.
\end{equation}
That is, the null energy condition requires that $RFF$ have a negative coefficient and so $2\alpha\mp\beta\geq0$, which coincides exactly with the WGC bound.

\subsection{Supersymmetry}
We can also ask what restrictions supersymmetry places on top of the two symmetries considered above. With $\mathcal{N}\geq2$ supersymmetry we expect there to be no correction to the charge-to-mass ratio of BPS states, much like was found for quantum corrections to the WGC in~\cite{Charles:2019qqt}. For the O($d,d;\mathbb{R}$) case this is easy to check: the heterotic string has $(\alpha,\beta)=(\frac{\alpha'}{16},-\frac{\alpha'}{8})$, so that $\beta=-2\alpha$ and the corrections are
\begin{equation}
	z_\text{ext} = 1 + \frac{4\alpha(1\mp 1)}{5p(p+q)} \,.
\end{equation}
The top sign corresponds to the choice in equation~\eqref{eq:Achoices} which places the gauge fields in the $\mathcal{N}=4$ gravity multiplet giving $z_\text{ext}=1$ as expected, while the lower sign places the gauge fields in vector multiplets giving positive corrections to $z_\text{ext}$.

For SL($2,\mathbb{R}$) we may gain some insight by using relations between scattering amplitudes for fields in the same supermultiplet. If $A_1$ is in the $\mathcal{N}=2$ gravity multiplet and $A_2$ is in the vector multiplet with $\phi,\theta$, then
\begin{equation}
	\mathcal{M}(h^+h^+h^-h^-) \sim \mathcal{M}(A_1^+A_1^+A_1^-A_1^-) \,.
\end{equation}
But the right-hand side has an $s^2$ term proportional to $\alpha_{1111}$ while the left-hand side has no $s^2$ term generated by the higher-derivative terms, and so $\alpha_{1111}$ must vanish. Similarly, for $\phi^\pm = \phi \pm i\theta$,
\begin{equation}
	\mathcal{M}(h^+h^-\phi^+\phi^-) \sim \mathcal{M}(A_1^+A_1^-\phi^+\phi^-)
\end{equation}
with now only the right-hand side having an $s^2$ term proportional to $\alpha_{11}$, so that it must be that $\alpha_{11}=0$. For the vector multiplet with $A_2$ and $\phi^\pm$, the relation
\begin{equation}
	4\alpha_2 s^2 + 8\alpha_1(t^2+u^2) = \mathcal{M}(\phi^+\phi^+\phi^-\phi^-) \sim \mathcal{M}(A_2^+A_2^+A_2^-A_2^-) = 8\alpha_{2222}s^2
\end{equation}
imposes $\alpha_1=0$. All told, only $\alpha_{2222}$, $\alpha_{22}$ and $\alpha_2$ are nonzero, giving positive correction to $z_\text{ext}$ only when fields in the vector multiplet are turned on.

\section{Discussion}
\label{sec:Disc}
The Einstein-Maxwell-dilaton-axion theory has a large number of possible four-derivative terms which correct the action in the effective action framework. The Wilson coefficients which control the relative sizes and signs of these terms may be partially constrained by appealing to scattering positivity bounds, but there remain allowed regions of the parameter space in which corrections to extremal black holes are at odds with the expectations of the WGC. Rather than working with a particular UV theory in order to determine more finely the form of the higher-derivative corrections, one may consider the implications of using symmetries inherited from the UV as an intermediate assumption. Our analysis holds whenever the dominant higher-derivative terms respect the symmetry in question.

The leading EMda action can have both an SL($2,\mathbb{R}$) and underlying O($d,d;\mathbb{R}$) symmetry. Imposing each of these individually on the action restricts the allowed higher-derivative terms to a small handful which then succumb to other considerations. We have shown that after having imposed SL($2,\mathbb{R}$) symmetry, positivity bounds are then enough to demonstrate the WGC in general. In imposing the O($d,d;\mathbb{R}$) symmetry we have found that the WGC requires a relationship between $\alpha$ and $\beta$ that follows from the null energy condition alone. Whether the null (or other) energy condition implies the mild form of the WGC in more general settings is a question worth further exploration.

Imposing SUSY on top of these two symmetries, we have found that the extremal charge-to-mass ratio is not corrected when black holes carry the gravity multiplet photon charge alone, as expected for BPS states.
Importantly, the (always) negative contributions from terms such as $\partial\phi\partial\phi FF$ are vital in canceling the positive contributions from other terms. It would be interesting, however, to consider what mileage one can get from considering the implications of SUSY alone on the four-derivative terms, again as an intermediate assumption on the UV theory. We leave this interesting question to future work.

\acknowledgments

We would like to thank Yuta Hamada and Yu-tin Huang for useful discussion. TN and GS thank Stefano Andriolo, Tzu-Chen Huang and Hirosi Ooguri for collaboration in a related topic~\cite{Andriolo:2020lul}.
The work of GL and GS is supported in part by the DOE grant DE-SC0017647 and the Kellett Award of the University of Wisconsin. TN is supported in part by JSPS KAKENHI Grant
Numbers JP17H02894 and JP20H01902. TN and GS gratefully acknowledge the hospitality of the Kavli Institute for Theoretical Physics (supported by NSF PHY-1748958) while part of this work was completed.

\appendix
\section{Positivity Bounds in Gravitational Theories}
\label{sec:grav_positivity}

In this appendix we review and and elaborate on the argument in Ref.~\cite{Hamada:2018dde} about positivity bounds in gravitational theories. For illustration, let us consider a scalar-graviton EFT up to four derivatives (generalization to more general EFTs is straightforward):
\begin{align}
	S=\int \d[4]{x}\,\sqrt{-g}\left[\frac{M_{\rm Pl}^2}{2}R-\frac{1}{2}\partial_\mu\phi\partial^\mu\phi+\alpha(\partial_\mu\phi\partial^\mu\phi)^2 +\cdots\right]\,,
\end{align}
where the dots stand for higher derivative terms. The IR expansion of scalar four-point amplitudes follows from this EFT as
\begin{align}
\label{IR_amplitude}
	\mathcal{M}(s,t)=\frac{1}{2M_{\rm Pl}^2}\frac{s^4+t^4+u^4}{stu}+2\alpha\left(s^2+t^2+u^2\right)+\cdots\,,
\end{align}
where $u=-(s+t)$. In particular the graviton $t$-channel exchange appearing in the first term behaves in the forward limit as $\displaystyle \sim -\frac{s^2}{M_{\rm Pl}^2 t}$, which is the main obstruction to deriving a positivity bound on the $s^2$ coefficient.

\medskip
To derive a rigorous bound on the $s^2$ coefficient, Ref.~\cite{Hamada:2018dde} employed an assumption that the graviton exchange is Reggeized by an infinite tower of higher spin states\footnote{The higher spin states are not necessarily particles appearing in tree-level exchange. They can also be multi-particle states appearing in loops, where angular momenta of relative motion play the role of spins.} and the full amplitudes is bounded as $<s^2$ in the Regge limit:
\begin{align}
\label{Regge_assumption}
|\mathcal{M}(s,t)|<s^2\quad\text{for large $s$ with $t<0$ fixed}.
\end{align}
Then they decomposed the amplitude as
\begin{align}
\label{decomposition}
\mathcal{M}(s,t)=\mathcal{M}_{\rm grav}(s,t)+\mathcal{M}_{\rm others}(s,t)\,,
\end{align}
where $\mathcal{M}_{\rm grav}$ is the graviton exchange Reggeized by the higher spin tower and $\mathcal{M_{\rm others}}$ is contributions from other states. More explicitly, the decomposition is performed such that the following conditions are satisfied:
\begin{enumerate}
\item The massless graviton pole appears in $\mathcal{M}_{\rm grav}$ only. As a result, $\mathcal{M}_{\rm others}$ has no massless poles in particular.
\item $\mathcal{M}_{\rm grav}$ is Reggezied by the higher spin states and bounded as $<s^2$ in the Regge limit:
\begin{align}
\label{grav_Regge_UV}
|\mathcal{M}_{\rm grav}(s,t)|<s^2\quad\text{for large $s$ with $t<0$ fixed}.
\end{align}
Combined with the assumption~\eqref{Regge_assumption}, this in turn guarantees that 
\begin{align}
\label{M_others}
|\mathcal{M}_{\rm others}(s,t)|<s^2\quad\text{for large $s$ with $t<0$ fixed}.
\end{align}
Here it is crucial to separate {\it Reggeized} graviton exchange satisfying the bound~\eqref{grav_Regge_UV}, rather than the simple graviton exchange. Otherwise, the subtracted amplitude $\mathcal{M}_{\rm others}$ would not satisfy the bound~\eqref{M_others} necessary for the following argument.

\item The Reggeization of the graviton exchange is controlled by a mass scale $M_{\rm Regge}$ of Regge states and its low-energy expansion is given schematically by
\begin{align}
\mathcal{M}_{\rm grav}=\frac{1}{2M_{\rm Pl}^2}\frac{s^4+t^4+u^4}{stu}+\sum_{n=2}^\infty\mathcal{O}(1)\times\frac{E^{2n}}{M_{\rm Pl}^2M_{\rm Regge}^{2n-2}}\,,
\end{align}
where $E^{2n}$ schematically denotes $s,t,u$-symmetric $n$-th order polynomials of $s,t,u$. Assuming that $M_{\rm Regge}$ is the only scale characterizing the Reggeization of graviton exchange, we assigned $\mathcal{O}(1)$ coefficients in the second term. Note that the sign of the $\mathcal{O}(1)$ coefficient of the $E^4$ term cannot be fixed at least by the present technology.
\end{enumerate}
Based on the assumption $1$, we may expand $\mathcal{M}_{\rm others}$ in the IR as
\begin{align}
\mathcal{M}_{\rm others}=\sum_{n,m=0}^\infty a_{n,m}s^nt^m\,.
\end{align}
Essentially because $\mathcal{M}_{\rm others}$ has no massless graviton poles and it is bounded as Eq.~\eqref{M_others}, positivity of $a_{2,0}$ follows under the standard assumptions of the positivity argument~\cite{Adams:2006sv}. Matching with Eq.~\eqref{IR_amplitude}, we find
\begin{align}
\label{alpha_decomposition}
\alpha=\frac{a_{2,0}}{2}+\mathcal{O}(1)\times\frac{1}{M_{\rm Pl}^2M_{\rm Regge}^2}
\end{align}
with positive $a_{2,0}$ and an $\mathcal{O}(1)$ coefficient with an unconstrained sign. Based on this, Ref.~\cite{Hamada:2018dde} concluded that positivity of the $s^2$ coefficient, $\alpha>0$, follows at least when the contribution from the gravitational Regge states, i.e., the second term in Eq.~\eqref{alpha_decomposition}, is subdominant. This is what we mean in the main text by {\it gravitational effects are subdominant}. Also one may rephrase this conclusion that $\alpha$ enjoys a bound\footnote{See Refs.~\cite{Alberte:2020jsk,Tokuda:2020mlf} for recent followup works on this generalized bound.}
\begin{align}
\alpha \geq \mathcal{O}(1)\times\frac{1}{M_{\rm Pl}^2M_{\rm Regge}^2}\,,
\end{align}
where the value of the $\mathcal{O}(1)$ coefficient depends on details of Reggeization and its sign cannot be fixed at least by the present technology. In the rest of this appendix, we provide some examples for such a decomposition.

\paragraph{Type II superstring.}

We have argued that the standard positivity $\alpha>0$ does not necessarily follow in the presence of gravity. Let us begin by providing an illustrative example for its violation. Consider the tree-level amplitude of an identical scalar originating from the 10D graviton in the type II superstring with a simple compactification on $T^6$:\footnote{See also Ref.~\cite{Tokuda:2020mlf} for more detailed study of its dispersion relation and the $s^2$ coefficient.}
\begin{align}
\label{type_II_amplitude}
\mathcal{M}_{\text{type-II}}(s,t)=-\frac{1}{2M_{\rm Pl}^2}\frac{s^4+t^4+u^4}{stu}\frac{\Gamma(-\alpha's/4)\Gamma(-\alpha't/4)\Gamma(-\alpha'u/4)}{\Gamma(\alpha's/4)\Gamma(\alpha't/4)\Gamma(\alpha'u/4)}\,,
\end{align}
where $\alpha'$ is the Regge slope. Its Regge limit is given by
\begin{align}
\mathcal{M}_{\text{type-II}}(s,t)\simeq \frac{1}{M_{\rm Pl}^2\,t}s^{2+\frac{1}{2}\alpha' t}
\end{align}
up to a phase factor and so the amplitude satisfies the condition~\eqref{grav_Regge_UV}. Then, one may decompose the amplitude in the form~\eqref{decomposition}, e.g., such that
\begin{align}
\mathcal{M}_{\rm grav}(s,t)=\mathcal{M}_{\text{type-II}}(s,t)\,,
\quad
\mathcal{M}_{\rm others}(s,t)=0\,.
\end{align}
Since there are no states which give the $s^2$ contribution bigger than $\displaystyle\frac{\mathcal{O}(1)}{M_{\rm Pl}^2M_{\rm Regge}^2}$, we cannot expect $\alpha>0$ in general. Indeed, the IR expansion of the amplitude~\eqref{type_II_amplitude} reads
\begin{align}
\mathcal{M}_{\text{type-II}}(s,t)=\frac{1}{2M_{\rm Pl}^2}\frac{s^4+t^4+u^4}{stu}+\mathcal{O}(E^8)\,,
\end{align}
where $\mathcal{O}(E^8)$ denotes the eighth and higher order derivatives. In particular, it has a vanishing four-derivative coefficient: $\alpha=0$, which provides an example for violation of the standard positivity $\alpha>0$. Note that a vanishing $s^2$ coefficient is required by $\mathcal{N}=8$ SUSY (in the 4D language) of the type II theories.

 \paragraph{Open string amplitudes.}

Next, as an illustrative example for theories with a positive $\alpha$, let us consider the case where the scalar originates from a 10D gauge boson in the open string spectrum. Then, the scalar four-point amplitude is given schematically as
\begin{align} 
\mathcal{M}_{\rm open}(s,t)=\mathcal{M}_{\rm disk}(s,t)+\mathcal{M}_{\rm annulus}(s,t)+\text{(higher genus amplitudes)}\,,
\end{align}
where the first two terms are the disk and annulus amplitudes, respectively. In the following we ignore higher genus amplitudes denoted by the third term. In the Regge limit with a negative $t$, each term is bounded as $<s^2$ and also the disk amplitude does not contain graviton exchange. Therefore, we may perform the decomposition~\eqref{decomposition} such that
\begin{align}
\mathcal{M}_{\rm grav}(s,t)=\mathcal{M}_{\rm annulus}(s,t)\,,
\quad
\mathcal{M}_{\rm others}(s,t)=\mathcal{M}_{\rm disk}(s,t)\,.
\end{align}
In particular, the $s^2$ coefficient originating from each amplitude is estimated as\footnote{We assumed that the scale of compactification and the volume of the cycles on which the brane wrapped are $\mathcal{O}(M_s)$. The same hierarchy is expected to hold as long as there is no unusual hierarchy between them.}
\begin{align}
\alpha|_{\rm grav}\sim\frac{1}{M_{\rm Pl}^2M_s^2}\,,
\quad
\alpha|_{\rm others}\sim\frac{1}{g_s}\frac{1}{M_{\rm Pl}^2M_s^2}
\end{align}
with $M_s$ and $g_s$ being the string scale and the closed string coupling,
so that the latter dominates over the former in the weakly coupled regime $g_s\ll1$. Since the latter has a positive $s^2$ coefficient, we conclude that
\begin{align}
\alpha\simeq \alpha_{\rm others}>0\,.
\end{align}
Note that in this example massive states generating a positive $s^2$ coefficient $\alpha_{\rm others}$ have the same mass scale as gravitational Regge states. The hierarchy $\alpha_{\rm others}\gg |\alpha_{\rm grav}|$ originates from the hierarchy $g_o\sim g_s^{1/2}\gg g_s$ between the open string coupling $g_o$ and the closed string coupling $g_s$. On the other hand, the same hierarchy $\alpha_{\rm others}\gg |\alpha_{\rm grav}|$ appears and thus the positivity of $\alpha$ follows also when there exists an intermediate state whose coupling is as small as the gravitational coupling, but whose mass is hierarchically lighter than gravitational Regge states. This is the case, e.g., when the dilaton is stabilized well below the string scale and also when the KK scale is well below the string scale and the KK graviton appears as an intermediate state.

\bibliography{Dualities_WGC_v2}

\end{document}